\documentclass[twoside,twocolumn]{article}

\usepackage{blindtext} % Package to generate dummy text throughout this template 

\usepackage[sc]{mathpazo} % Use the Palatino font
\usepackage[T1]{fontenc} % Use 8-bit encoding that has 256 glyphs
\usepackage{microtype} % Slightly tweak font spacing for aesthetics

\usepackage[english]{babel} % Language hyphenation and typographical rules

\usepackage[tmargin=1.5in, lmargin=0.75in, rmargin=0.75in, bmargin=1.0in]{geometry}
\usepackage[hang, small,labelfont=bf,up,textfont=it,up]{caption} % Custom captions under/above floats in tables or figures
\usepackage{booktabs} % Horizontal rules in tables

\usepackage{lettrine} % The lettrine is the first enlarged letter at the beginning of the text

\usepackage{enumitem} % Customized lists
\setlist[itemize]{noitemsep} % Make itemize lists more compact
\usepackage{amsfonts,amssymb,amsmath,amsgen,amsopn,amsbsy,theorem,graphicx,epsfig}

\usepackage{abstract} % Allows abstract customization
\usepackage{titlesec} % Allows customization of titles
\renewcommand\thesection{\Roman{section}} % Roman numerals for the sections
\renewcommand\thesubsection{\roman{subsection}} % roman numerals for subsections
\titleformat{\section}[block]{\large\scshape\centering}{\thesection.}{1em}{} % Change the look of the section titles
\titleformat{\subsection}[block]{\large}{\thesubsection.}{1em}{} % Change the look of the section titles

\usepackage{fancyhdr} % Headers and footers
\usepackage{titling} % Customizing the title section

\usepackage{hyperref} % For hyperlinks in the PDF

\pretitle{\begin{center}\Huge\bfseries} % Article title formatting
\posttitle{\end{center}} % Article title closing formatting
\title{When deep learning meets security} % Article title
\author{%
\textsc{Majd Latah}\thanks{Corresponding Author} \\[1ex] % Your name
\normalsize Department of Computer Science \\ % Your institution
\normalsize Ozyegin University \\ % Your institution
\normalsize \href{mailto:latahmajd@gmail.com}{latahmajd@gmail.com} % Your email address
}
\date{} % Leave empty to omit a date
\renewcommand{\maketitlehookd}{%
\begin{abstract}
{\fontsize{11}{20}\selectfont Deep learning is an emerging research field that has proven its effectiveness towards deploying more efficient intelligent systems. Security, on the other hand, is one of the most essential issues in modern communication systems. Recently many papers have shown that using deep learning models can achieve promising results when applied to the security domain. In this work, we provide an overview for the recent studies that apply deep learning techniques to the field of security.
}

\end{abstract}
}

\begin{document}

% Print the title
\maketitle

%----------------------------------------------------------------------------------------
%	ARTICLE CONTENTS
%----------------------------------------------------------------------------------------

\section{Introduction}
Machine learning techniques have proven to be very useful in networking [1] in general and security related topics in particular [2,3,4]. Deep learning [5-7], on the other hand, has improved the state-of-the-art for many machine learning tasks such as speech recognition, objection detection and natural language understanding. The main advantage is that this approach allows learning very complex functions by a general-purpose learning approach [5]. It is worth noting that we exclude deep intrusion detection systems from this study due to the fact that many other researchers have already conducted a similar survey to cover that particular subject [8,9].

\section{Deep learning in security}

\subsection{Malware Detection}

Tobiyama et al. [10] proposed malware process detection based on process behavior. The authors used long short term memory (LSTM) for feature extraction and convolutional neural network (CNN) for classification. A process behavior is a sequence of API calls. The features were extracted from the process behavior log files which were transferred to an image that contains local features. Theses local features mostly represents the process activities. Therefore, one can apply CNN in order to capture these local features and correctly classify these images. An overview of the proposed method in [10] is shown in Figure 1. In the experimental study the authors used 81 malware process log files and 69 benign process log files for the training and validation stages. The authors executed the malware files in the Cuckoo Sandbox and traced the process behavior in order to determine the produced and injected processes. The authors also validated the classifier with 5-fold cross validation. The best result (AUC= 0.96) achieved when the feature image size was 30x30.

\begin{figure}[ht!]
\centering
\includegraphics[width=80mm]{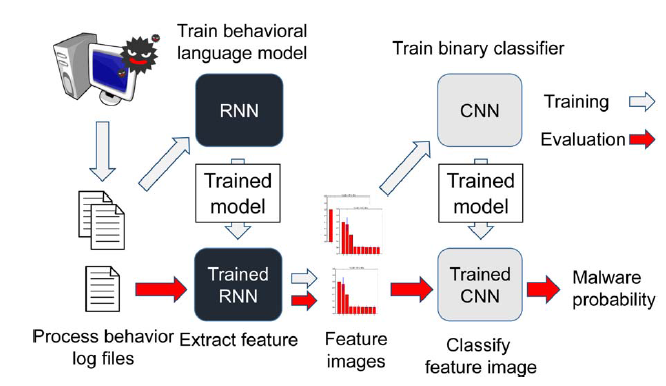}
\caption{Overview of the proposed deep learning approach in [10] \label{overflow}}
\end{figure}

Rhode et al. [11] investigated the possibility of prediction whether an executable is malicious or not based on behavioural data. The study showed that the model achieved high level of accuracy (94\%) based only on the first 5 seconds of the file execution using 3000 ransomware samples without prior exposure to these samples. In addition, the model achieved an accuracy of 96\% in less than 10 seconds. The selected features were 10 continuous machine activity data metrics instead of using categorical API calls. Due to the fact that API calls can be manipulated which may lead to incorrect classification for the input samples. In addition, continous data allows a large number of states to be represented in a small vector. As shown in Figure 2, Cuckoo Sandbox was used in order to collect activity data of Portable Executable (PE) samples. The collected features are: system CPU usage, user CPU use, packets sent, packets received, bytes sent, bytes received, memory use, swap use, the total number of processes currently running and the maximum process ID assigned. The authors used gated recurrent unit (GRU) in stead of LSTM cells due to their training speed. The authors stated that the model should be re-trained regularly with newly discovered samples, which may lead to adjustments in the proposed architecture too.

\begin{figure}[ht!]
\centering
\includegraphics[width=80mm]{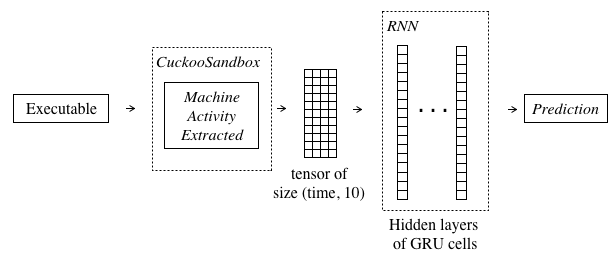}
\caption{Overview of the proposed deep learning model in [11] \label{overflow}}
\end{figure}

Chen et al. [12] proposed, HeNet, a hierarchical ensemble neural network based on control flow characterization of program execution. As shown in Figure 3, HeNet consists of a low-level behavior model and a top-level ensemble model. HeNet was tested against ROP attacks against Adobe Reader 9.3 in Windows R7 32 bit. The dataset used in the experimental study consists of 348 benign and 299 malicious PDF samples. HeNet achieved high level of accuracy (100\%) and zero level of false positives (0\%). In addition to a higher classification accuracy when compared to traditional machine learning algorithms.

\begin{figure}[ht!]
\centering
\includegraphics[width=80mm]{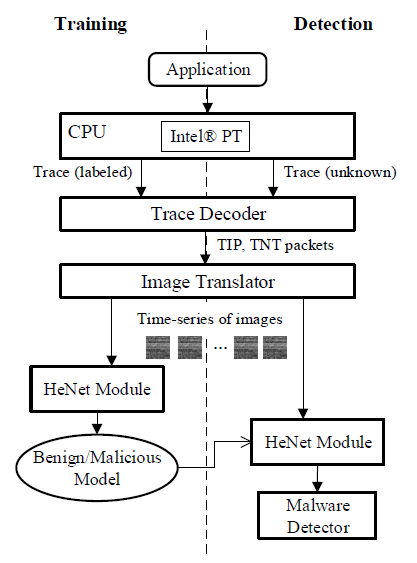}
\caption{Overview of the proposed deep learning model in [12] \label{overflow}}
\end{figure}

Hardy et al. [13] proposed a malware detection model based on stacked AutoEncoders (SAEs). The model uses Windows API calls produced from the collected PE files. As shown on Figure 4, the PE parser is used to extract the Windows API calls from each file. The API query database, converts the API calls to a 32-bit representation for the corresponding API functions. Thereafter, the SAE is used in order to perform feature learning, fine-tuning and malware detection. The experimental study conducted on large dataset collected from Comodo Cloud Security Center. The dataset contained 50000 file samples, where 22500 are malware, 22500 are benign files, and 5000 are unknown. The proposed model outperformed ANN, SVM, NB, and DT in malware detection with 96.85\% level of accuracy.

\begin{figure}[ht!]
\centering
\includegraphics[width=80mm]{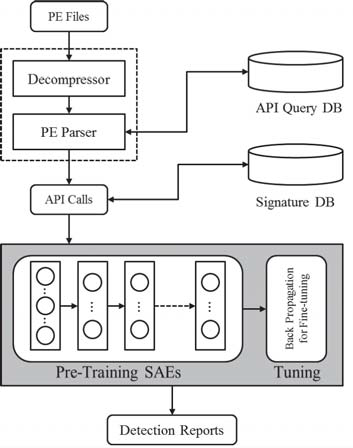}
\caption{Overview of the proposed deep learning model in [13] \label{overflow}}
\end{figure}

Hou et al. [14], proposed an Android malware detection framework based on Linux kernel system calls and stacked AutoEncoders (SAEs). A new dynamic analysis method named Component Travelsal was introduced for automatically execution of the code routines of each given Android app. In order to capture the relationships among the system calls, a weighted directed graph was constructed where each graph node represents a system call and its size indicates its frequency, whereas a directed edge indicates the sequential flow of system calls made and includes a weight that implies the frequency the successor node called after the predecessor node. Each node with its weight, each edge with its weight, as well as the in-degree and out-degree of each of the nodes is used as the input feature for the proposed deep learning model. The experimental study was conducted on large dataset collected from Comodo Cloud Security Center. The dataset contains 3000 android apps, half of which are benign, while the other half are malicious. Compared with traditional machine learning methods, the detection performance is enhanced by using deep learning framework with a high level of accuracy (93.68\%).

\begin{figure}[ht!]
\centering
\includegraphics[width=80mm]{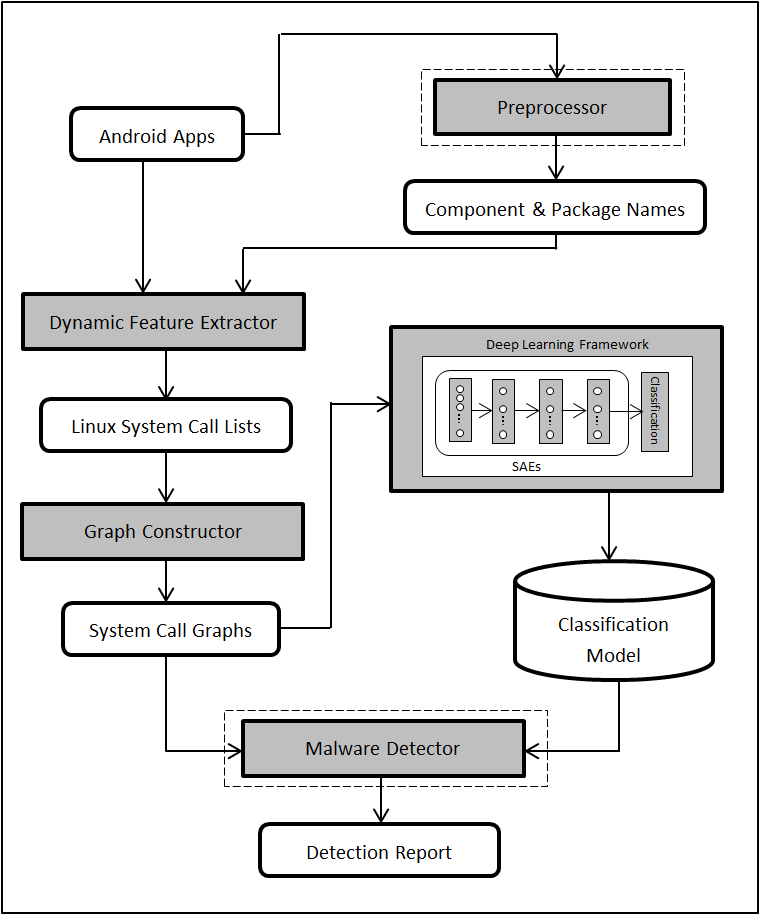}
\caption{Overview of the proposed deep learning model in [14] \label{overflow}}
\end{figure}

Yuan et al. [15], proposed, DroidDetector, an online deep-learning based Android malware detection engine. The authors conducted static and dynamic analyses to extract features from each app. The extracted features fall into three main categories: (1) required permissions, (2) sensitive APIs and (3) dynamic behaviors. DroidDetector achieved 96.76\% detection accuracy, which outperforms traditional machine learning techniques. In the static phase includes parsing the two files AndroidManifest.xml and classes.dex in order to obtain a total of 120 permissions required by the app. The dynamic phase includes running each app in DroidBox in order to execute a dynamic taint analysis and monitor a total of 13 of the app actions. As shown in Figure 6, the deep learning model used in this study consists of two phases, the unsupervised pretraining phase and supervised back-propagation phases. the pre-training phase, the Deep Belief Networks (DBN) is hierarchically built by stacking a number of Restricted Boltzmann Machines (RBM). In the back-propagation phase, the pre-trained DBN was tuned with labeled samples in a supervised way.  The experimental study was conducted on three public app sets. The first benign app set was randomly crawled from the Google Play Store and contains a total of 20000 apps. The other two malicious app sets were respectively collected from the Contagio Community (there are only about 400 apps) and Genome Project (including 1260 malicious apps).

\begin{figure}[ht!]
\centering
\includegraphics[width=80mm]{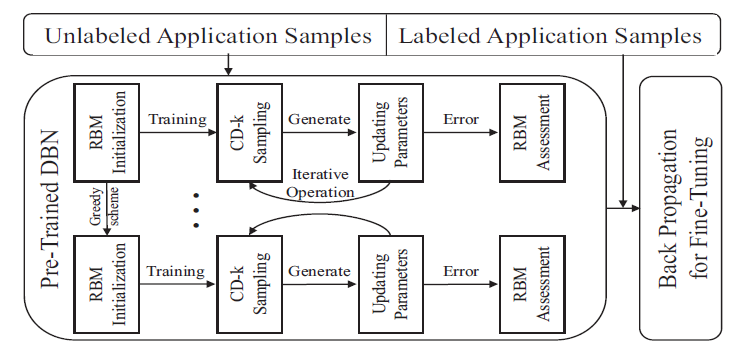}
\caption{Overview of the proposed deep learning model in [15] \label{overflow}}
\end{figure}

Huang and Stokes [16] proposed, MtNet, a multi-task deep learning malware classification architecture which is trained for two tasks including binary classification which predicts whether an unknown file is malicious or benign and 100-class family classification which predicts if the file belongs to one of 98 important families, a generic malware class, or the benign class. The authors used low-level features extracted from dynamic analysis of the file as input for the training stage. These features are a sequence of application programming interface (API) call events plus their parameters and a sequence of null-terminated objects recovered from system memory during emulation. The final number of selected features was reduced to 50000 based on mutual information feature selection. Training a neural network with this large input dimension is computationally intensive. Therefore, the authors used random projection technique in order to reduce the data size. MtNet trained and tested on an extremely large dataset consisting of 6.5 million, 2.85 million examples were extracted from malicious files and 3.65 million from benign files. The set of malicious files has 1.3 million belonging to the 98 malware families and 1.55 million from the generic malware class. A set of 4.5 million samples were used for training and another set of 2.0 million for a hold out test set. The experimental results showed that MtNet achieved a binary malware error rate of 0.358\% and family error rate of 2.94\%. The multi-task learning improved the classification results and showed low false positive rates (under 0.07\%).

\begin{figure}[ht!]
\centering
\includegraphics[width=80mm]{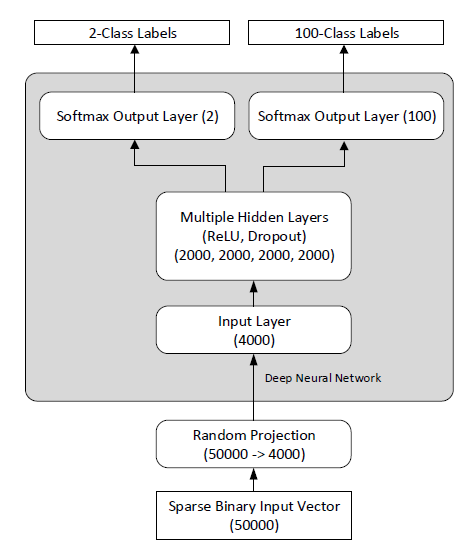}
\caption{Overview of the proposed multi-task deep learning malware classification architecture in [16] \label{overflow}}
\end{figure}

Azmoodeh et al. [17], proposed a deep Eigenspace learning approach to classify malicious and bening IoT applications. They extracted OpCode sequence of 1078 benign and 128 malware from ARM compatible IoT applications. The selected features (OpCodes) of each sample are converted into a graph which is for classification based on deep convolutional networks. They used Objdump in order to extract the OpCodes. Thereafter, one can use n-gram Op-Code sequence to classify malware using their disassembled codes. In addition, they proposed class-wise information gain to overcome the problem of imbalanced datasets and select the top 82 features. As shown in Figure 8, the proposed approach consists of 2 phases: (1) OpCode-Sequence Graph Generation phase and (2) Deep Eigensapce Learning phase. Eigensapce learning was proposed due to the fact that Eigenvectors and eigenvalue are two main components in the graph's spectrum. The first two eigenvectors and eigenvalues of the samples are used as input values for the model. The proposed system achieved an accuracy of 98.37\% and a precision rate of 98.59\%. In addition to, the ability to mitigate junk code insertion attacks.

\begin{figure}[ht!]
\centering
\includegraphics[width=80mm]{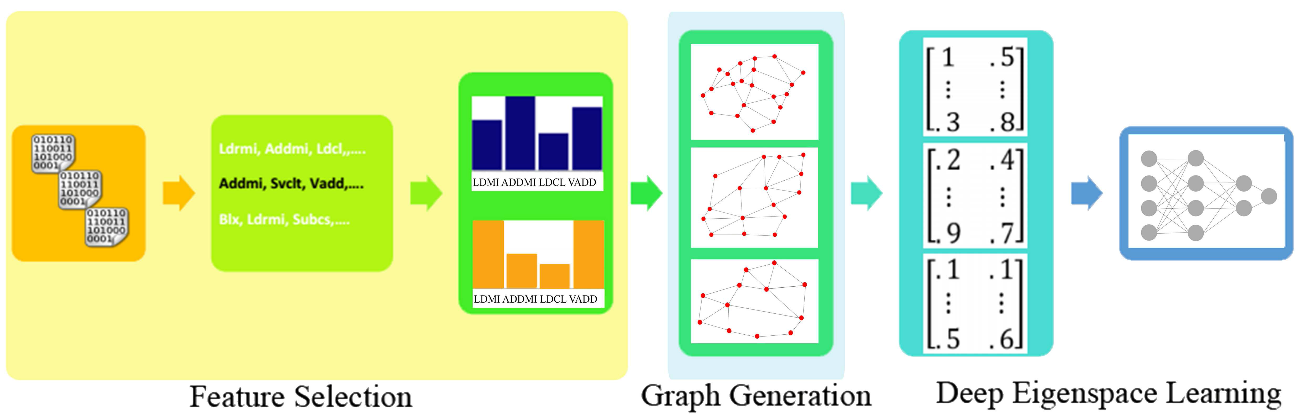}
\caption{Overview of the proposed method in [17] \label{overflow}}
\end{figure}

Kolosnjaji et al. [18], proposed a classification of malware system call sequences based on convolutional and recurrent network layers. As shown in Figure 9, the authors combined convolutional and recurrent layers in one neural network in which the convolutional layer is used for feature extraction. The input of the system is 60 distinct system calls. The dataset used in study was collected from three different sources: Virus Share, Maltrieve and private collections. Sample labels are obtained using services of VirusTotal where the hash of the malware file compared to the service's database. Then the authors performed clustering on the signatures from different antivirus programs in order to obtain ground truth classes from the resulting clusters where these clusters contain 4753 malware samples. The experimental study showed that the proposed approach outperformed the traditional machine learning approaches and achieved an average of 85.6\% on precision and 89.4\% on recall. 

\begin{figure}[ht!]
\centering
\includegraphics[width=80mm]{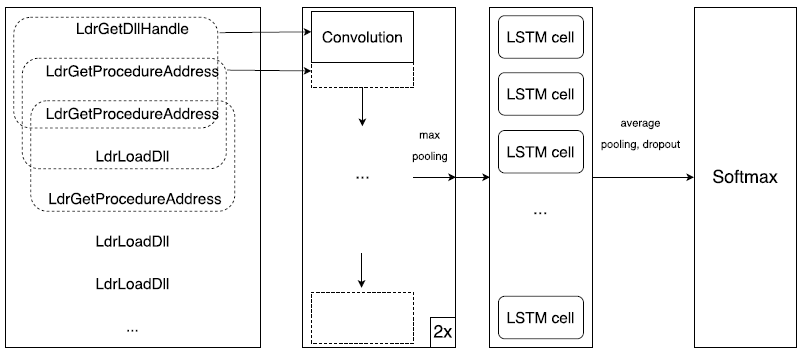}
\caption{Overview of the proposed deep learning model in [18] \label{overflow}}
\end{figure}

Rigaki and Garcia [19], showed that malware behavioral patterns can be modified using GANs in order to mimic Facebook chat.  The real-life testing scenario was developed using the Stratosphere behavioral IPS in a router, whereas the malware and the GAN were deployed in the local network, and the command \& control server was deployed in the cloud. The authors used an open source Remote Access Trojan (RAT), called Flu, which was modified in order to receive the input from the generator and adapt its behavior accordingly. Flu inquires the GAN using the HTTP API, which in turn reply with three parameters: total byte size, duration of the next network flow, and time delta between the current flow and the next one.

\subsection{Botnet Detection}
Botshark [20] is a deep learning-based approach for botnet traffic analyzing from both centralized and P2P botnets. Botshark consists of two deep structures where the first layer employs stacked Autoencoders for feature extraction and the second layer uses CNNs in order to train a classifier for botnet detection. The authors used network traffic from ISCX dataset [21] which includes 44.97\% malicious flows from 16 different centralized and decentralized botnet topologies as well as normal traffic. The authors extracted NetFlows from network traffic using the Argus [22] tool. The extracted features contain: (1) byte-based features, (2) time-based features, (3) packet-based features. The experimental study showed 0.91\% true positive rate with 0.13\% of false positive rate.

The authors of [23] used LSTM approach to model the behavior of network traffic as a sequence of states that changes over time. The authors used LSTM approach to model the behavior of network traffic as a sequence of states that changes over time. The behavior of a connection is computed based on three features of each flow: size, duration and periodicity. Then, assigning to each flow a state symbol based on the extracted features and an assignment strategy. Finally, each connection will have its own string of symbol used for represents its behavior. The proposed LSTM architecture was evaluated against two different datasets where the first one was used for training and the second one was used for testing purpose. The authors used sampling approach for dealing with the imbalanced datasets. They also studied the optimal length of connection states required for the input layer in LSTM and consequently the best results were achieved by 25 connection states. The LSTM model showed 0.809\% attack detection rate with 0.03\% of false alarm rate when tested on the second dataset. The experimental results showed that this approach was able to detect TCP based malicious connections. However, it failed in identifying most of the HTTP and HTTPS traffic. In addition to some of SMTP (SPAM) traffic. A solution for this problem, is observe the network traffic across multiple layers of the network and monitor the sequences of bot activities together which possibly can be merged based on a context and aim [24]. 

In [25], Yin et al. used generative adversarial networks (GANs) for generating 'fake' samples in order to expand the number of labeled samples where a 3-layer LSTM network was used as generator and the discriminator was replaced with a botnet detector. The authors selected 16 flow-based features. The experimental study was conducted on ISCX botnet dataset. The false positive rate of the GAN-based model was decreased from 19.19\% to 15.59\%. 

Tran et al. [26] proposed a novel LSTM approach in order to detect modern sophisticated botnets that employ domain generation algorithm (DGA) to generate a large number of domains that can be used in order to communicated with Command and Control (C\&C) server. DGA classification can be seen as a multiclass task [25], which can be either retrospective or real-time manner. The real time detection which is mainly based on the domain name and linguistic features. This is much difficult due to the fact that linguistic features it can be bypassed by the malware author [27]. In general, LSTM is sensitive to the multiclass imbalance problem. In other words, it is naturally biased towards the prevalent classes, which results in an inability of detecting the uncommon other DGA families [26]. The new proposed algorithm adopts the cost-sensitivity principle [28] in order to target the class imbalance problem, by making the learning biased towards the small classes. The authors modified the backward pass (i.e. the error computation) of the original LSTM. This new algorithm has showed an enhancement of 7\% in terms of macro-averaging recall, precision and F1-score with respect to the original LSTM and other state-of-the-art solutions. The cost-sensitive LSTM, however, reduced the accuracy on the prevalent non-DGA class. It is also worth mentioning that cost-sensitive LSTM is superior to RUSBoost, oversampling and Threshold-moving methods and achieved much higher macro-averaging F1-score with respect to HMM, C5.0, LSTM, the cost-sensitive SVM, cost-sensitive C4.5 and Weighted Extreme Learning Machine on the multiclass imbalanced dataset [26].

\begin{figure}[ht!]
\centering
\includegraphics[width=80mm]{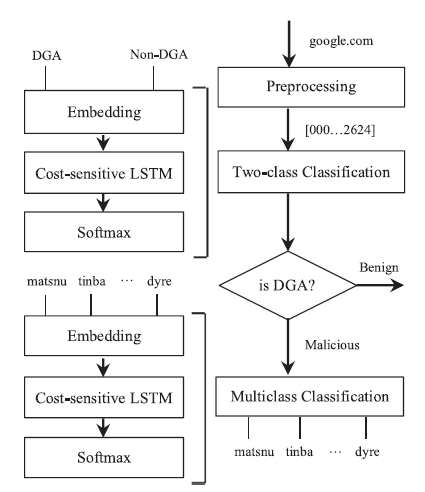}
\caption{Overview of the proposed method in [26] \label{overflow}}
\end{figure}

On the other hand, in the context of social bot detection for social media platforms, traditional deep learning techniques for text classification depended mainly on only textual features [28,29]. However, employing additional features can be more efficient and shows better results. In [29] deep neural network based on contextual (LSTM) architecture that exploits both content and metadata in order to detect social bots at the tweet level. Contextual LSTM is a natural language processing (NLP) model based on deep learning paradigm. The dataset [30] includes over 8386 accounts and 11834866 tweets. The authors used both 10 account-level features for the first level and 6 tweet level features for the second level. This approach showed a high level of accuracy (> 99\% AUC) for user-level detection.

\subsection{Malicious Code Detection}
Hendler et al. [31] used deep learning model for detecting malicious PowerShell commands based on character-level convolutional neural networks. The authors used a large dataset which consists of 66388 distinct PowerShell commands; 6290 labeled as malicious and 60098 labelled as benign. The authors treated the command as a raw signal at character level and applied to it a one dimensional CNN for text classification task. The input feature length is 1,024, therefore if a command is longer than that it will be truncated. The best performance was achieved by an ensemble detector that combines a traditional NLP-based classifier with a CNN-based classifier.

\begin{figure*}[ht!]
\centering
\includegraphics[width=160mm]{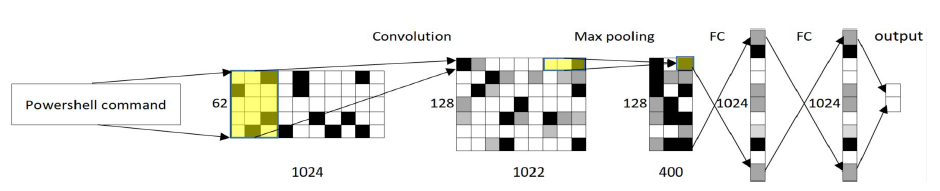}
\caption{Overview of the proposed CNN architecture in [31] \label{overflow}}
\end{figure*}

Wang et al. [32], used stacked denoising auto-encoders in order to extract high-level features from JavaScript code. Denoising auto-encoder is an extension of traditional auto-encoders, which can reconstruct the inputs from a noisy corrupted data by minimizing the reconstruction loss. The JavaScript code was converted to binary feature vectors. As shown in Figure 12, a logistic regression classifier is applied to the output of the last hidden layer of the auto-encoder in order to classify the JavaScript code to the appropriate class. The authors evaluated our model on a large dataset, which contains over 27000 samples. For dimensionality reduction the authors used sparse random projection method to reduce target dimensionality which resulted in 480 dimensionalities. The experimental study showed that the proposed model can attain a high level of accuracy (95\%) with a false positive rate less than 4.2\%. The limitation of the model proposed in [32] is the long training time. However, it has a high-speed testing time. In addition, the proposed method outperformed other traditional machine learning algorithms in terms of precision, recall, f-measure, RMSE and testing time.

\begin{figure}[ht!]
\centering
\includegraphics[width=80mm]{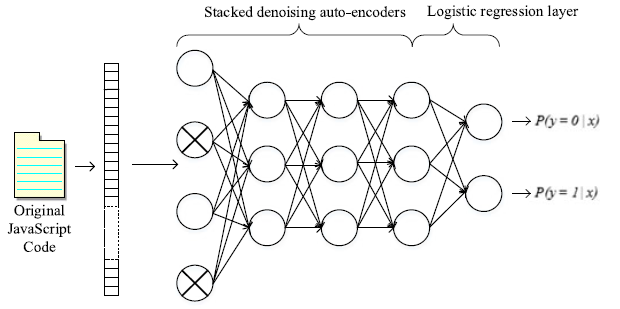}
\caption{Overview of the proposed approach in [32] \label{overflow}}
\end{figure}

Saxe and Berlin [33] proposed, eXpose, a convolutional neural network approach for detecting malicious URLs, file paths and registry keys. The training dataset used in this research consists of a total of 14788254 instances: 8332360 benign and 6455894 malicious samples whereas the testing dataset consists of a total of 11531955 instances: 10121981 benign and 1409974 malicious samples. The authors used n-gram feature extractor and manually extracted features. These features were randomly hashed into 1024-dimensional vector. Thereafter, they were fed into a deepMLP model. eXpose outperformed manual feature extraction approaches, achieving a 5\%-10\% detection rate gain at 0.1\% false positive rate compared to these baselines. 

\section{Conclusion}
This paper gives an overview of the potential of deep learning techniques in the security field. We mainly observed the high accuracy and efficiency of deep learning models in malware and botnet detection. In addition to their application in malicious code detection. These models showed a significant improvement when compared to tradition machine learning approaches. Therefore, we conclude that these models will have an increased adoption in modern security solutions and applications.

%----------------------------------------------------------------------------------------
%	REFERENCE LIST
%----------------------------------------------------------------------------------------

%----------------------------------------------------------------------------------------

\end{document}